\documentclass[10pt]{iopart}
\usepackage{iopams}

\usepackage{graphicx}

\usepackage{bm}
\usepackage[unicode=true,
 bookmarks=true,bookmarksnumbered=false,bookmarksopen=false,
 breaklinks=false,pdfborder={0 0 0},pdfborderstyle={},backref=false,colorlinks=true]
 {hyperref}
\hypersetup{pdfborderstyle={},pdfborderstyle={},pdfborderstyle={},pdfborderstyle={},pdfborderstyle={},pdfborderstyle={},linkcolor=blue,citecolor=blue}

\newcommand\beq{\begin{equation}}
\newcommand\eeq{\end{equation}}
\newcommand\beqa{\begin{eqnarray}}
\newcommand\eeqa{\end{eqnarray}}
\newcommand{\text}[1]{\mbox{#1}}

\def\bal#1\eal{\begin{align}#1\end{align}}

\newcommand{\barr}{\left\{\begin{array}{l}}
\newcommand{\earr}{\end{array}\right.}

\newcommand{\tw}{\text{\scriptsize{TW}}}
\newcommand{\ff}{v^{\tw}_\gamma(\xi)}

\begin{document}

\title[Laplace transform in quantum mechanics]{Revisiting the Laplace transform in quantum mechanics: correcting a flawed approach for the stationary Schr\"odinger equation}

\author{Luis M~B\'aez and Andr\'es Santos}

\address{Departamento de F\'isica, Universidad de Extremadura, E-06006 Badajoz, Spain}
\ead{andres@unex.es}
\vspace{10pt}
\begin{indented}
\item[]\today
\end{indented}

\begin{abstract}
The Laplace transform is a valuable tool in physics, particularly in solving differential equations with initial or boundary conditions. A 2014 study by Tsaur and Wang (2014 \emph{Eur.~J.~Phys.} \textbf{35} 015006) introduced a Laplace-transform-based method to solve the stationary Schr\"odinger equation for various potentials. However, their approach contains critical methodological flaws: the authors disregard essential boundary conditions and apply the residue theorem incorrectly in the inverse transformation process. These errors ultimately cancel out, leading to correct results despite a flawed derivation. In this paper, we revisit the use of the Laplace transform for the one-dimensional Schr\"odinger equation, clarifying correct practices in handling boundary conditions and singularities. This analysis offers a sound and consistent framework for the application of Laplace transforms in stationary quantum mechanics, underscoring their educational utility in quantum mechanics coursework.

\end{abstract}

%
\vspace{2pc}

\section{Introduction}

The Schr\"odinger equation is a fundamental cornerstone of non-relativistic quantum mechanics, governing the quantum behavior of systems at the microscopic scale. The eigenfunctions of the Hamiltonian operator, which represent physical solutions to the stationary Schr\"odinger equation, are central to understanding quantum phenomena. Its one-dimensional form holds particular significance, serving as a crucial tool in both academic research and practical applications.

Laplace transforms are crucial in physics as they provide a powerful tool for solving differential equations that describe a wide variety of physical systems.
 When applied to the stationary Schr\"odinger equation, this technique can streamline the solution process and uncover key properties of quantum systems that may be difficult to discern using conventional methods.

About a decade ago, Tsaur and Wang (TW) employed the Laplace-transform method to derive exact solutions to the stationary one-dimensional Schr\"odinger equation for various potential functions $U(x)$~\cite{TW14}. The primary objective of this paper is to expose critical errors in TW's methodology, including the omission of essential boundary conditions and the incorrect application of the residue theorem.

\section{Some properties of the Laplace transform}

Given an exponentially bounded function $f(x)$ defined in the domain $x>0$, its Laplace transform is defined as $F(s)=\mathcal{L}[f(x)]=\int_0^\infty \rmd x\, \rme^{-s x}f(x)$, where $s$ is a complex variable~\cite{S99}.
This integral converges if $f(x)$ grows no faster than an exponential function as $x\to\infty$.

Among several key properties of the Laplace transform, three are particularly relevant here. First, for derivatives of
$f(x)$:
\begin{equation}
\label{1}
\mathcal{L}[f'(x)]=sF(s)-f(0),\quad \mathcal{L}[f''(x)]=s^2F(s)-sf(0)-f'(0).
\end{equation}
Next, the inverse Laplace transform is (Mellin's inverse formula)
\begin{equation}
\label{2}
f(x)=\mathcal{L}^{-1}[F(s)]=\lim_{\gamma\to\infty} f_\gamma(x),\quad  f_\gamma(x)=\frac{1}{2\pi \rmi}\int_{a-\rmi\gamma}^{a+\rmi\gamma}\rmd s\, \rme^{s x}F(s),
\end{equation}
where $a$ is chosen to be greater than the real part of all singularities of $F(s)$.
If the product $\rme^{s x}F(s)$ decays sufficiently fast on a left semicircle supported by the vertical line $s=a$ (Bromwich contour), the residue theorem can be applied to evaluate the inverse Laplace transform. This gives
\begin{equation}
\label{3}
f(x)=\sum_{j}\text{Res}\left[\rme^{s x}F(s)\right]_{s_j},
\end{equation}
where $\{s_j\}$ are the poles of $\rme^{s x}F(s)$.

As the third property, if the function $f(x)$ has finite moments, defined as
$M_p=\int_0^\infty \rmd x\, x^p f(x)$,
then these moments are directly connected to the series expansion of $F(s)$ in powers of $s$:
\begin{equation}
\label{4}
F(s)=\sum_{p=0}^\infty (-1)^p M_p\frac{s^p}{p!}.
\end{equation}
This expression illustrates how the coefficients in the power series of $F(s)$ correspond to the moments $M_p$ of the original function $f(x)$.

\section{Application to the stationary one-dimensional Schr\"odinger equation}

The stationary one-dimensional Schr\"odinger equation for a certain potential $U(x)$ reads
\begin{equation}
\label{SE}
-\frac{\hbar^2}{2m}\psi''(x)+U(x)\psi(x)=E\psi(x),
\end{equation}
where $\hbar$ is Planck's constant, $m$ is the particle's mass, $E$ represents the energy level, and $\psi(x)$ denotes the wavefunction, which is generally a complex-valued function..
When applied  to \Eref{SE}, the main steps of the Laplace-transform method are:
\begin{enumerate}
    \item
    \label{step1}
  Introduce scaled variables $x\to\xi$, $\psi(x)\to v(\xi)$.
  \item
  \label{step2}
 Derive the second-order differential equation for $v(\xi)$  from \Eref{SE}.
  \item
  \label{step4}
  Apply the Laplace transform to obtain a differential equation for $V(s)=\mathcal{L}[v(\xi)]$.
  \item
  \label{step5}
  Solve for the physical solution $V(s)$, obtaining a quantization condition along the process.
  \item
  \label{step6}
  Invert the Laplace transform to recover $v(\xi)=\mathcal{L}^{-1}[V(s)]$.
  \item
  \label{step7}
  Return to the original wavefunction $v(\xi)\to\psi(x)$.
\end{enumerate}

Steps (\ref{step1}) and (\ref{step2}) align with conventional \cite{ER85} and Fourier-transform  \cite{M98,P04,E06} solution methods, whereas steps (\ref{step4})--(\ref{step7}) are unique to the Laplace-transform approach.
Note that the application of the Laplace transform to $v(\xi)$ requires knowledge of this function only for $\xi \geq 0$ to fully determine $\psi(x)$. This reduction of the domain is possible under two key circumstances: either the potential $U(x)$ is symmetric, ensuring that its bound-state wavefunctions possess well-defined parity (even or odd), or the entire domain of the spatial variable $ x$ is naturally mapped into the region $\xi \geq 0$.

\begin{table}
\caption{\label{tab0}Expressions for the potential $U(x)$, the dimensionless energy parameter, and the scaled variables $\xi$ and $v(\xi)$ for three prototypical oscillators. In the cases of the Morse and the modified P\"oschl--Teller oscillators, the strengths of the potentials are characterized by the dimensionless parameters  $c={\sqrt{2mA}}/{\alpha\hbar}$ and  $\ell=\frac{1}{2}\left(\sqrt{1+{8mA}/{\alpha^2\hbar^2}}-1\right)$, respectively.}
\footnotesize
\begin{tabular}{@{}lllll}
\br
Oscillator&$U(x)$&Energy parameter&$\xi$&$ v(\xi)$\\
\mr
Harmonic
&$\displaystyle{\frac{1}{2}m\omega^2x^2}$&$\displaystyle{n=\frac{E}{\hbar\omega}-\frac{1}{2}}$&$
\displaystyle{\sqrt{\frac{m\omega}{\hbar}}x}
$&$\rme^{\xi^2/2}\psi(x)$\\
Morse&$
A\left(\rme^{-2\alpha x}-2\rme^{-\alpha x}\right)
$
& $\displaystyle{n=c-\frac{1}{2}-\frac{\sqrt{-2mE}}{\alpha\hbar}}$&$2c\rme^{-\alpha x}$&$\xi^{n-c+\frac{1}{2}}\psi(x)$\\
\hspace{-0.3cm}
$
\begin{array}{l}
\text{Modified}\\ \text{P\"oschl--Teller}
\end{array}
$
&$
-A\text{sech}^2(\alpha x)$&$\displaystyle{\mu=\frac{\sqrt{-2mE}}{\alpha\hbar}}$&
$
\displaystyle{\text{tanh}(\alpha x)}$&
$
\left(1-\xi^2\right)^{-\mu/2}\psi(x)
$\\
\br
\end{tabular}\\
\end{table}
\normalsize

\Tref{tab0} lists three representative potentials, along with the corresponding variable transformations  $(x, \psi) \to (\xi, v)$ of step (\ref{step1}) and the dimensionless parameter ($n$ or $\mu$) characterizing the energy level. The resulting second-order differential equations for $v(\xi)$, step (\ref{step2}), are shown in the second column of \Tref{tab1}.
The other quantum problems discussed in \cite{TW14} reduce, after a suitable change of variables, to either one of the equations in \Tref{tab0} or to the Bessel equation.

\begin{table}
\caption{\label{tab1}Differential equations for $v(\xi)$ and $V(s)$, along with the quantization conditions, for the oscillators in \Tref{tab0}. In the third column, $v_0\equiv v(0)$ and $v_0'\equiv v'(0)$}
\footnotesize
\begin{tabular}{@{}llll}
\br
Oscillator&Equation for $v(\xi)$&Equation for $V(s)$&
\hspace{-0.3cm}
$
\begin{array}{l}
\text{Quantization}\\ \text{condition}
\end{array}
$\\
\mr
Harmonic
&$v''-2\xi v'+2nv=0$&$
\hspace{-0.1cm}
\barr
\hspace{-0.2cm}2sV'+(s^2+2n+2)V\\
\hspace{-0.2cm}-v_0s-v_0'=0
\earr
$&$n=0,1,2,\ldots$\\
Morse&$
\hspace{-0.1cm}
\barr
\hspace{-0.2cm}\xi v''+2(c-n)v'\\
\hspace{-0.2cm}+\left(c-\frac{\xi}{4}\right)v=0
\earr
$&$
\hspace{-0.1cm}
\barr
\hspace{-0.2cm}\left(\frac{1}{4}-s^2\right)V'\\
\hspace{-0.2cm}+\left[c+2(c-n-1)s\right]V\\
\hspace{-0.2cm}-[2(c-n)-1]v_0=0
\earr
$
&
$
\hspace{-0.1cm}
\barr
\hspace{-0.2cm}n=0,1,2,\ldots\\
\hspace{-0.2cm}\left(n<c-\frac{1}{2}\right)
\earr
$\\
\hspace{-0.3cm}
$
\begin{array}{l}
\text{Modified}\\ \text{P\"oschl--Teller}
\end{array}
$
&$
\hspace{-0.1cm}
\barr
\hspace{-0.2cm}(1-\xi^2)v''-2(\mu+1)\xi v'\\
\hspace{-0.2cm}+[\ell(\ell+1)-\mu(\mu+1)]v=0
\earr
$&
$
\hspace{-0.1cm}
\barr
\hspace{-0.2cm}-s^2 V''+2(\mu-1)sV'\\
\hspace{-0.2cm}+[(\ell+\mu)(\ell+1-\mu)+s^2]V\\
\hspace{-0.2cm}-v_0s-v_0'=0
\earr
$&
$
\ell-\mu=0,1,2\ldots
$\\
\br
\end{tabular}\\
\end{table}
\normalsize

Following step (\ref{step4}), the differential equations for $V(s)$ corresponding to the three prototypical oscillators are presented in the third column of \Tref{tab1}.
Note that, while $\xi$ is positive-definite and only $v_0$ enters the differential equation for $V(s)$ in the Morse oscillator case, the harmonic and modified P\"oschl--Teller oscillators involve both $v_0$ and $v_0'$ due to the symmetric nature of their potentials.

As an illustration of steps (\ref{step5})--(\ref{step7}), let us consider the harmonic oscillator. The physical solution $v(\xi)$ must be regular at $\xi = 0$, allowing it to be expressed as a power series: $v(\xi) = v_0 + v_0'\xi + \sum_{k=2}^\infty v_k \xi^k$. Consequently, the Laplace transform $V(s)$ takes the form
$V(s) = \sum_{k=0}^\infty V_k s^{-(k+1)}$,
where $V_0 = v_0$, $V_1 = v_0'$, and $V_k = v_k k!$ for $k \geq 2$.
Substituting this expansion of $V(s)$ into its corresponding first-order differential equation yields the recurrence relation
\begin{equation}
V_{k+2} = -2(n - k)V_k,\quad k\geq 0.
\end{equation}
For a generic value of the energy parameter $n$, the recurrence relation leads to the asymptotic behaviors $V_{k+2}/V_k \sim 2k$ and $v_{k+2}/v_k \sim 2/k$ for large $k$. Consequently, the series expansion $V(s) = \sum_{k=0}^\infty V_k s^{-(k+1)}$ diverges, indicating that $V(s)$ has an essential singularity at infinity. Furthermore, the solution $v(\xi)$ exhibits asymptotic growth of the form $v(\xi) \sim \rme^{\xi^2}$ for large $\xi$. This behavior translates into the asymptotic form $\psi(x) \sim \rme^{\xi^2/2}$, which is unphysical for bound-state wavefunctions.

On the other hand, this singular behavior is avoided if $n$ is a nonnegative  integer and either $v_0'$ vanishes (for $n$ even) or $v_0$ equals zero (for $n$ odd). In either case, $V(s)$ has a polynomial form:
\begin{equation}
V(s)\propto \sum_{k=0}^{\lfloor n/2\rfloor}\frac{ (-1)^k }{2^{2k}k!}s^{-(n-2k)-1},
\end{equation}
where $\lfloor n/2\rfloor$ denotes the floor function of $n/2$. This concludes step (\ref{step5}). Applying the inverse Laplace transform to $V(s)$ directly gives $v(\xi) \propto H_n(\xi)$, where $H_n(\xi)$ represents the Hermite polynomials. Consequently, the physical wavefunctions are $\psi(x) \propto e^{-\xi^2/2} H_n(\xi)$, with the proportionality constant determined by the normalization condition.

The Morse and P\"oschl--Teller potentials are handled in an analogous manner. The fourth column of \Tref{tab1} lists the respective quantization conditions derived by imposing the requirement for physical solutions.
Those physical solutions $v(\xi)$ and $V(s)=\mathcal{L}[v(\xi)]$ for the ground state and the first excited state are presented in \Tref{tab2}, where the normalization constants are omitted for simplicity.

Let us illustrate the property in  \Eref{4}  with the first excited state of the Morse potential, for which $v(\xi) = (2c - 2 - \xi)e^{-\xi/2}$. Its moments are $M_p = \int_0^\infty \mathrm{d}\xi\, \xi^p v(\xi) = 2^{p+2} p! (c - 2 - p)$. Substituting into \Eref{4} gives
\begin{equation}
V(s) = 4 \sum_{p=0}^\infty (c - 2 - p)(-2s)^p = 4 \left[\frac{c-2}{1+2s} + \frac{2s}{(1+2s)^2}\right],
\end{equation}
where we have used the mathematical identities $\sum_{p=0}^\infty x^p = 1/(1 - x)$ and $\sum_{p=0}^\infty px^p = x/(1 - x)^2$. It is straightforward to verify that this expression for $V(s)$ matches the one provided in \Tref{tab2}.

For the harmonic and P\"oschl--Teller oscillators, the functions $v(\xi)$ are polynomials, leading to divergent moments. Consequently, their Laplace transforms $V(s)$ cannot be expressed as series expansions in positive powers of $s$.

\begin{table}
\caption{\label{tab2}Solutions for the ground state and  first excited state corresponding to the oscillators in \Tref{tab0}. The fifth column presents the expressions for the functions $V^{\tw}(s)$ as derived in \cite{TW14}, obtained under the assumption that $v_0 = v_0' = 0$. Here, $k_\ell(s)$ is the modified spherical Bessel function of the second kind~\cite{AS72,Weisstein}.
}
\footnotesize
\begin{tabular}{@{}lllll}
\br
Oscillator&
$
\hspace{-0.3cm}
\begin{array}{l}
\text{Quantum}\\ \text{number}
\end{array}
$
&$v(\xi)$&$V(s)$&$V^{\tw}(s)$\\
\mr
Harmonic&
$n=0$&$1$&$s^{-1}$&$\rme^{-s^2/4}s^{-1}$\\
&$n=1$&$\xi$&$s^{-2}$&$\rme^{-s^2/4}s^{-2}$\\
Morse&$n=0$&$\rme^{-\xi/2}$&$(s+\frac{1}{2})^{-1}$&
$
\hspace{-0.1cm}
\barr
\hspace{-0.2cm}(s-\frac{1}{2})^{2c-1}\\
\hspace{-0.2cm}\times(s+\frac{1}{2})^{-1}
\earr
$\\
& $
\hspace{-0.1cm}
\barr
\hspace{-0.2cm}
n=1\\
\hspace{-0.2cm}\left(\text{if } c>\frac{3}{2}\right)
\earr
$
&$
\hspace{-0.1cm}
\barr
\hspace{-0.2cm}\left(2c-2-\xi\right)\\
\hspace{-0.2cm}\times \rme^{-\xi/2}
\earr
$&$
\hspace{-0.1cm}
\barr
\hspace{-0.2cm}(2c-2)(s+\frac{1}{2})^{-1}\\
\hspace{-0.2cm}-(s+\frac{1}{2})^{-2}
\earr$
&$
\hspace{-0.1cm}
\barr
\hspace{-0.2cm}(s-\frac{1}{2})^{2c-2}\\
\hspace{-0.2cm}\times(s+\frac{1}{2})^{-2}
\earr
$\\
\hspace{-0.3cm}
$
\begin{array}{l}
\text{Modified}\\ \text{P\"oschl--Teller}
\end{array}
$
&$\mu=\ell$&$1$&$s^{-1}$&$s^{\ell}k_\ell(s)
$\\
&$\mu=\ell-1$&$\xi$&$s^{-2}$&$s^{\ell-1}k_\ell(s)
$\\
\br
\end{tabular}\\
\end{table}
\normalsize

\section{Critique of TW's approach}

Although TW's paper \cite{TW14} holds significant pedagogical value, there are two notable flaws in their derivations, specifically in steps (\ref{step4}) and (\ref{step6}) above.

In progressing to step (\ref{step4}), TW applies \Eref{1} without accounting for the initial (or boundary) values $v_0$ and $v_0'$. They justify this by writing \cite{TW14} `\emph{For simplicity, in section 3 we ignore all the initial conditions $f^{(k)}(0)$ for $k =0, 1, 2, \ldots , n - 1$ [\ldots]}' and `\emph{The initial conditions can be ignored because all the Schr\"odinger equations considered in
this paper are linear and homogeneous.}' However, this argument is flawed. Ignoring the initial conditions $v_0$ and $v_0'$ is not merely a matter of convenience but of consistency; if the final result $v(\xi)$ contradicts the assumption $v_0=v_0'=0$, then neglecting these conditions in step (\ref{step4}) is unjustified.
Moreover, while the Schr\"odinger equation itself is indeed homogeneous, its Laplace transform is generally inhomogeneous due to the presence of $v_0$ and $v_0'$.
Among the various cases examined in \cite{TW14}, the neglect of $v_0$ is only justified in the specific case of the radial equation for a two-dimensional free particle, where the Bessel equation arises.

The flaw in step (\ref{step6}) arises from using \Eref{3} when the conditions necessary for transitioning from \Eref{2} to \Eref{3} are not met. For example, in the case of the harmonic oscillator, TW assume that $\mathcal{L}^{-1}[s^k]=0$ with $k=0,1,\ldots$ because $s^k$ does not have any poles. However, \Eref{3} is not applicable when $F(s)=s^k$ with $k\geq 0$. In particular, $\mathcal{L}^{-1}[1]=\delta(x)$.
Interestingly, the two flaws ultimately cancel each other out, resulting in the correct function $v(\xi)$.
In this context, it is important to note that while Chen's approach for solving the Morse potential~\cite{C04} contains similar flaws, the methods employed by Pimentel and de Castro for the harmonic potential~\cite{PC13}, as well as Englefield's seminal contributions to the subject~\cite{E68}, do not exhibit these inconsistencies.

From \Tref{tab2} we see that $(v_0,v_0')=(1,0)$ and $(0,1)$ for the ground and first excited states, respectively, in the cases of the harmonic and modified Poschl--Teller oscillators. Moreover, in the Morse oscillator, $(v_0,v_0')=(1,-\frac{1}{2})$ and $(2c-2,-c)$ for the ground and first excited states, respectively. In each case, the function $V(s)$ satisfies the inhomogeneous differential equation listed in \Tref{1}, using the appropriate values of $v_0$ and $v_0'$. On the other hand, the function $V^{\tw}(s)$, as derived in \cite{TW14} and included in \Tref{tab2}, adheres to the same differential equation only under the restrictive condition $v_0 = v_0' = 0$.

As a simple example, consider  the ground state of the harmonic oscillator. How do TW derive $v(\xi)=1$ from $V^{\tw}(s)=\rme^{-s^2/4}s^{-1}$? They do so by expanding $V^{\tw}(s)$ into a series:
\begin{equation}
V^{\tw}(s) = s^{-1} + \sum_{k=1}^\infty \frac{s^{2k-1}}{(-4)^k k!}.
\end{equation}
They then claim, as previously noted, that $\mathcal{L}[s^{2k-1}] = 0$ for $k \geq 1$ invoking the residue theorem, as indicated by \Eref{3}. However,  this assumption is fundamentally flawed.
In fact, we demonstrate in the subsequent analysis that the inverse Laplace transform $v^{\tw}(\xi) = \mathcal{L}^{-1}[V^{\tw}(s)]$ of $V^{\tw}(s) = \rme^{-s^2/4}s^{-1}$ results in a function characterized by extreme singularity.

\begin{figure}
\centering
      \includegraphics[width=0.45\columnwidth]{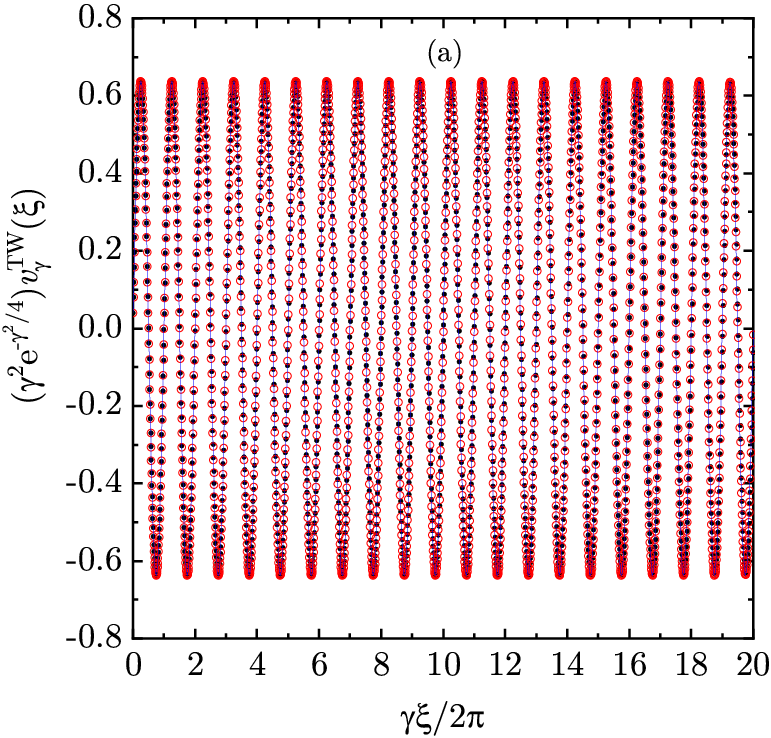}\hspace{0.5cm}\includegraphics[width=0.45\columnwidth]{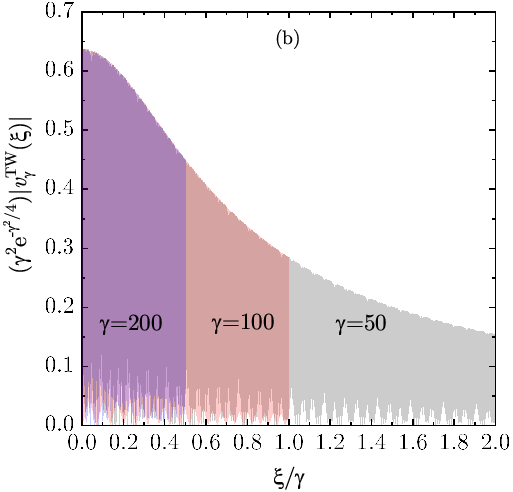}
      \caption{(a) Plot of $\gamma^{2}\rme^{-\gamma^2/4}\ff$ vs $\gamma\xi/2\pi$ for $\gamma=50$ (dots), $100$ (open circles), and $200$ (solid line). (b) Plot of $\gamma^{2}\rme^{-\gamma^2/4}|\ff|$ vs $\xi/\gamma$ for $\gamma=50$ (up to $\xi/\gamma=2$), $100$ (up to $\xi/\gamma=1$), and $200$ (up to $\xi/\gamma=0.5$).
  \label{fig1}}
\end{figure}

According to \Eref{2}, $v^{\tw}(\xi)=\lim_{\gamma\to\infty}v^{\tw}_\gamma(\xi)$, where $\ff=(2\pi \rmi)^{-1}\int_{a-\rmi\gamma}^{a+\rmi\gamma}\rmd s\, \rme^{s x}V^{\tw}(s)$ and $a>0$.
After the change of variable $s=a+\rmi y$, $\ff$ can be rewritten as
\begin{equation}
\fl
\ff=\frac{\rme^{a \xi}}{\pi}\int_{0}^{\gamma}\rmd y\, \rme^{(y^2-a^2)/4}\frac{a\cos[y(\xi-a/2)]+y\sin[y(\xi-a/2)]}{a^2+y^2}.
\end{equation}
Since in the limit $\gamma\to\infty$ the value of $a$ becomes irrelevant, we can take the limit $a\to 0^+$, yielding
\begin{equation}
\ff=\frac{1}{\pi}\int_{0}^{\gamma}\rmd y\, \rme^{y^2/4}y^{-1}\sin(y\xi).
\end{equation}
Numerical evaluation  reveals that $\ff \approx \rme^{\gamma^2/4} \gamma^{-2} A_\gamma(\xi) \sin(\gamma \xi)$, where the envelope function $A_\gamma(\xi)$ exhibits two distinct regimes. For $\xi \sim \gamma^{-1}$, $A_\gamma(\xi)$ approaches a constant value $A_\gamma \simeq \frac{191}{300}$. This behavior is illustrated in \Fref{fig1}(a), where the rescaled function $\gamma^{2}\rme^{-\gamma^2/4}\ff\sim\sin(\gamma\xi)$ is plotted against $\gamma \xi / 2\pi$ for three large $\gamma$ values. As shown, there is excellent mutual agreement, with the oscillation wavelength being approximately $1$.
For $\xi \sim 1$ and $\xi \sim \gamma$, the envelope of the oscillations decreases with $\xi/\gamma$, as illustrated in \Fref{fig1}(b). In this regime, the plot of $\gamma^{2}\rme^{-\gamma^2/4}|\ff|$ versus $\xi/\gamma$ closely follows the envelope function, aside from high-frequency oscillations that superimpose on it.

In summary, the function $\ff$ oscillates very rapidly (with a wavelength of $2\pi\gamma^{-1}$) and an enormous amplitude on the order of $\rme^{\gamma^2/4}\gamma^{-2}$, which  decays slowly over a scale of approximately $\gamma$.
All of this underscores the highly singular nature of $v^{\tw}(\xi) = \lim_{\gamma \to \infty} v^{\tw}_\gamma(\xi)$ and the pathological character of the Laplace transform $V^{\tw}(s) = \rme^{-s^2/4}s^{-1}$ derived under the condition $v_0 = 0$, in stark contrast to the true functions $v(\xi) = 1$ and $V(s) = s^{-1}$.

An alternative way to characterize $v^{\tw}(\xi)$ is through the moments $M_p=\int_0^\infty\rmd \xi\,\xi^p \left[v^{\tw}(\xi)-1\right]$ of the difference $v^{\tw}(\xi)-1=\mathcal{L}^{-1}[s^{-1}(\rme^{-s^2/4}-1)]$. From \Eref{4} we have
\begin{equation}
M_p=
\left\{
\begin{array}{ll}
  0&\text{if $p=$even},\\
  \frac{(-1)^{(p-1)/2}p!!}{(p+1) 2^{(p+1)/2}}&\text{if $p=$odd}.
\end{array}
\right.
\end{equation}
As observed, the even moments of the deviation $v^{\tw}(\xi)-1$ vanish identically, reflecting symmetry properties, while the odd moments alternate in sign and grow in magnitude at an extremely rapid rate as the order increases. This behavior emphasizes the highly oscillatory and divergent nature of the moments associated with \( v^{\tw}(\xi) \), reinforcing its departure from regular physical functions.

\section{Concluding remarks}

In conclusion, while TW's method yields correct final results for the quantum systems under consideration, the inconsistencies in their application of the Laplace transform and its inverse highlight critical shortcomings in their approach. Specifically, neglecting essential initial and boundary conditions, as well as mismanaging singularities in the Laplace-domain representations, undermines the mathematical rigor and reliability of their derivations. These issues stress the importance of a more meticulous and systematic treatment of such conditions to ensure the validity of results and to avoid potentially flawed or misleading conclusions in the analysis of quantum systems.

\ack
AS acknowledges financial support from Grant No.~PID2020-112936GB-I00 funded by MCIN/AEI/10.13039/501100011033.

\section*{Data availability statement}
All data that support the findings of this study are included within the article (and any supplementary files).

\section*{References}

\begin{thebibliography}{10}
\expandafter\ifx\csname url\endcsname\relax
  \def\url#1{{\tt #1}}\fi
\expandafter\ifx\csname urlprefix\endcsname\relax\def\urlprefix{URL }\fi
\providecommand{\eprint}[2][]{\url{#2}}

\bibitem{TW14}
Tsaur G and Wang J 2014 {\em Eur. J. Phys.\/} {\bf 35} 015006

\bibitem{S99}
Schiff J~L 1999 {\em The {L}aplace {T}ransform.{T}heory and {A}pplications\/}
  Undergraduate {T}exts in {M}athematics (New York, NY: Springer)

\bibitem{ER85}
Eisberg R and Resnick R 1985 {\em Quantum Physics of Atoms, Molecules, Solids,
  Nuclei, and Particles\/} 2nd ed (New York: Wiley)

\bibitem{M98}
Mu{\~n}oz G 1998 {\em Am. J. Phys.\/} {\bf 66} 254--256

\bibitem{P04}
Ponomarenko S~A 2004 {\em Am. J. Phys.\/} {\bf 72} 1259--1260

\bibitem{E06}
Engel A 2006 {\em Am. J. Phys.\/} {\bf 74} 837

\bibitem{AS72}
Abramowitz M and Stegun I~A (eds) 1972 {\em {Handbook of Mathematical
  Functions}\/} (New York: Dover) pp.~ 443--445

\bibitem{Weisstein}
Weisstein E~W 2004 Modified {S}pherical {B}essel {F}unction of the {S}econd
  {K}ind
  \url{https://mathworld.wolfram.com/ModifiedSphericalBesselFunctionoftheSecondKind.html}

\bibitem{C04}
Chen G 2004 {\em Phys. Lett. A\/} {\bf 326} 55--57

\bibitem{PC13}
Pimentel D~R~M and {de Castro} A~S 2013 {\em Eur. J. Phys.\/} {\bf 34} 199--204

\bibitem{E68}
Englefield M~J 1968 {\em J. Aust. Math. Soc.\/} {\bf 8} 557--567

\end{thebibliography}
\providecommand{\newblock}{}

\end{document}